\documentclass{sig-alternate-10pt}
\newcommand{\comment}[1]{}

\usepackage{amsmath}
\usepackage{amssymb}
\usepackage{balance}

\usepackage{array}
\usepackage{algorithm}
\usepackage{algpseudocode}
\usepackage{threeparttable}
\usepackage{multirow}
\usepackage{subfigure}

\newcommand{\power}{\ensuremath{\mbox{\sc Power}}}
\newcommand{\yeast}{\ensuremath{\mbox{\sc Yeast}}}

\newcommand{\grqc}{\ensuremath{\mbox{\sc GrQc}}}
\newcommand{\hepth}{\ensuremath{\mbox{\sc HepTh}}}
\newcommand{\oregon}{\ensuremath{\mbox{\sc Oregon}}}

\begin{document}

\title{Classifying Latent Infection States \\in Complex Networks \vspace{-0.6cm}}
%
% You need the command \numberofauthors to handle the 'placement
% and alignment' of the authors beneath the title.
%
% For aesthetic reasons, we recommend 'three authors at a time'
% i.e. three 'name/affiliation blocks' be placed beneath the title.
%
% NOTE: You are NOT restricted in how many 'rows' of
% "name/affiliations" may appear. We just ask that you restrict
% the number of 'columns' to three.
%
% Because of the available 'opening page real-estate'
% we ask you to refrain from putting more than six authors
% (two rows with three columns) beneath the article title.
% More than six makes the first-page appear very cluttered indeed.
%
% Use the \alignauthor commands to handle the names
% and affiliations for an 'aesthetic maximum' of six authors.
% Add names, affiliations, addresses for
% the seventh etc. author(s) as the argument for the
% \additionalauthors command.
% These 'additional authors' will be output/set for you
% without further effort on your part as the last section in
% the body of your article BEFORE References or any Appendices.

\numberofauthors{3} %  in this sample file, there are a *total*
% of EIGHT authors. SIX appear on the 'first-page' (for formatting
% reasons) and the remaining two appear in the \additionalauthors section.
%
\author{
% You can go ahead and credit any number of authors here,
% e.g. one 'row of three' or two rows (consisting of one row of three
% and a second row of one, two or three).
%
% The command \alignauthor (no curly braces needed) should
% precede each author name, affiliation/snail-mail address and
% e-mail address. Additionally, tag each line of
% affiliation/address with \affaddr, and tag the
% e-mail address with \email.
%
% 1st. author
\alignauthor
Yeon-sup Lim\\
       \affaddr{University of Massachusetts Amherst}\\
       \email{ylim@cs.umass.edu}
\alignauthor
Bruno Ribeiro\\
       \affaddr{Carnegie Mellon University}\\
       \email{ribeiro@cs.cmu.edu}
\alignauthor
Don Towsley\\
       \affaddr{University of Massachusetts Amherst}\\
       \email{towsley@cs.umass.edu}
}
\maketitle

\begin{abstract}
%In this paper, we study the problem of a malfunction (e.g., computer viruses) spread in networks. Our concern is how to identify the infected nodes without individually inspecting all nodes in networks. Given the partial nodes that are observed (inspected), we propose the measure, Infection Betweenness, for identifying the status of unobserved nodes assuming that a malfunction is propagating into a network from a single source with the SI model. We then suggest the Geodesic Infection Betweenness which has a lower computational overhead than the Infection Betweenness. We empirically investigate the estimation performance based on our measures comparing with simple heuristics; we find that the Infection Betweeneess is a good indicator to estimate the status of unobserved nodes. Compared with Infection Betweenness, the Geodesic Infection Betweenness itself has a limitation to provide enough information for correctly estimating the status of unobserved nodes; however, we find that the performance close to the Infection Betweenness can be obtained by using machine-learning algorithms with the Geodesic Infection Betweenness and other node features such as degree.
Algorithms for identifying the infection states of nodes in a network are crucial for understanding and containing infections.
Often, however, only a relatively small set of nodes have a known infection state.
Moreover, the length of time that each node has been infected is also unknown.
This missing data -- infection state of most nodes and infection time of the unobserved infected nodes -- poses a challenge to the study of real-world cascades.

In this work, we develop techniques to identify the latent infected nodes in the presence of missing infection time-and-state data.
Based on the likely epidemic paths predicted by the simple susceptible-infected epidemic model, we propose a measure (Infection Betweenness) for uncovering these unknown infection states. Our experimental results using machine learning algorithms show that Infection Betweenness is the most effective feature for identifying latent infected nodes.

\end{abstract} 

% A category with the (minimum) three required fields
%\category{H.4}{Information Systems Applications}{Miscellaneous}
%A category including the fourth, optional field follows...
%\category{D.2.8}{Software Engineering}{Metrics}[complexity measures, performance measures]

%\terms{Network Management}

%\keywords{ACM proceedings, \LaTeX, text tagging} % NOT required for Proceedings
\vspace{-0.2cm}

\section{Introduction}
Networks are underlying mediums for the spread of epidemics such as diseases, rumors, and computer viruses. Determining the infection state of nodes is the first step to taking corrective or preventive action to stop or slow the spread of an epidemic. Unfortunately, the infection state of nodes is often unknown; for example: in the spread of computer malware (say, a contaminated email attachment) over a large organization, IT specialist will likely only inspect the computers of users that open trouble tickets; a similar problem occurs with the spread of rumors over online social networks. Hence, the problem of effectively identifying the infection state of unobserved nodes given a set of observed nodes is of central importance in the study of infection cascades.

In this work we consider a network where an epidemic starts from a single source.
%In Section~\ref{sec:conclusion} we discuss possible extensions to the multiple source case.
Each node appears in one of two states:(i) susceptible, capable of being infected, (ii) infected, able to spread the epidemic further. We also assume that the infection state of a subset of nodes is known and the full network structure (adjacency matrix) is available.
Our research question is: {\em Given a set of nodes with known infection state and the network topology can we correctly uncover the unknown infection state of the remaining nodes?}

The contributions of this work are the following:
\begin{itemize}
    \item We introduce a measure for estimating the state of unobserved nodes, denoted Infection Betweenness. Our simulation results using simple infection models show that our measure-based method classified nodes with an accuracy of up to $90\%$ while it finds up to $80\%$ of infected nodes in a network.
    \item We investigate the impact of network characteristics, such as the degree distribution and clustering coefficient, on the estimation performance of our approach. Our observations indicate that machine learning algorithms using our measure gets more accurate as the degree distribution becomes less positively-skewed and has a smaller standard deviation.
\end{itemize}

The remainder of the paper is organized as follows: Section \ref{sec:problem} depicts the problem statement. Section \ref{sec:measure} introduces Infection Betweenness. Section \ref{sec:results} represents the experimental result about the performance of Infection Betweenness with machine learning algorithms. Section \ref{sec:relatedwork} reviews the related literature. Finally, Section \ref{sec:conclusion} presents our conclusions and future work.

\vspace{-0.1cm}
\section{Problem Statement} \label{sec:problem}

%\subsection{Problem Statement}
Let $G(V, E)$ be an undirected graph where $V$ is a set of nodes and $E \subseteq V^2$ is a set of edges. Suppose that an epidemic starts at a single node (denoted ``source'') and propagates to neighbors in $G(V, E)$. Let $X_i \in \{ 0, 1\}$ denotes to the state of node $i \in V$ where $X_i=0$ means node $i$ is susceptible and $X_i=1$ that it is infected. Assume that an infected node contaminates neighbors at rate $\lambda$. Then,
\begin{eqnarray*}
    X_i &:& 0 \rightarrow 1 \quad \text{at rate } \lambda \sum_{j \in n(i)} X_j,
\end{eqnarray*} 
where $n(i)$ is the set of neighborhood of $i$.
%In the SI model, each node occupied one of two states, susceptible and infected, and the infection propagation time between nodes is exponentially distributed with homogeneous rate $\lambda$.

Assume that there are $l$ nodes with observed infection state $L = \{(1, X_1)$, ..., $(l, X_l))\}$. There are also $u = |V| - l$ nodes with unknown infection state, $U =\{x_{l+1}$ ..., $x_{l+u}\}$; $l$ is typically much smaller than $u$.
Given the set of observed nodes $L$ and the adjacency matrix $\mathbf{A}$ of the network, our goal is to correctly assign an infection state $X_i$ to node $i = l+1,\ldots,l+u$. 
\comment{
Table \ref{tab:notation} presents a table of notations used throughout this work.

\begin{table}[h!!!]
\vspace{-0.1cm}
\small
\centering
\caption{Table of Notation}
\label{tab:notation}
\begin{tabular}{cl}
  \hline
  Symbol & Explanation \\
  \hline
  \hline
  $V$ & Set of nodes in a graph \\
  $E$ & Set of edges in a graph \\
  $L$ & Set of nodes with observed infection state \\
  $U$ & Set of nodes with unknown infection state \\
  $I_{o}$ & Set of observed infected nodes \\
  $S_{o}$ & Set of observed susceptible nodes \\
  \hline
\end{tabular}
\vspace{-0.1cm}
\end{table}
}
  \begin{table*}[!ht]
  \small
    \centering
    \caption{Topologies }
    \label{tab:topology}
    \begin{center}
    \begin{threeparttable}[b]
    \begin{tabular}{ c c c c c c c c m{7cm} }
        \hline
         Topology & Type & $n$  & $m$ & $c$ & $\sigma$ & $s$ & $d$  \footnotemark[2]& {\centering Description} \\
        \hline
        \hline
%        \jazz{} & 198 & 1742 & 0.6175 & 6 & Jazz musician network obtained from The Red Hot Jazz Archive digital database \cite{gleiser03} \tabularnewline
%        \lesmis{} & 77 & 254 & 0.5731 & 5 & Coappearance network of characters in the novel Les Miserables \cite{Knuth} \tabularnewline
 %       \dolphin{} & 62 & 159 & 0.2590 & 8 & Social Network of frequent associations between 62 dolphins in a community living off Doubtful Sound, New Zealand \cite{Lusseau}\tabularnewline
%        \umass{} & 41 & 121 & 0.2342 & 5 & Collaboration Network at the department of computer science in the University of Massachusetts Amherst\tabularnewline
%        \adjnoun{} & 112 & 425 & 0.1728 & 5 & adjacency network of common adjectives and nouns in the novel David Copperfield by Charles Dickens \cite{Newman}\tabularnewline
        \yeast{} & Biological & 1870 & 2277 & 0.0672 & 3.1374 & 6.5044 & 19 & Yeast Protein Interaction Network \cite{Yeast} \tabularnewline
        \grqc{} & Collaboration & 5242 & 28980 & 0.5296 & 7.9179 & 3.8317 & 17 & Collaboration networks from ArXiv General Relativity and Quantum Cosmology \cite{collaboration}\tabularnewline
        \hepth{} & Collaboration& 9877 & 51971 & 0.4714 & 6.1864 & 3.0213 & 18 & Collaboration networks from ArXiv High Energy Physics \cite{collaboration} \tabularnewline
        \power{} & Device & 4941 & 6594 & 0.0801 & 1.7913 & 2.1898 & 46 & Topology of the Western States Power Grid of the United States \cite{Power} \tabularnewline
        \oregon{} & Device & 11174 & 23409 & 0.2964 & 33.0948 & 46.4017 & 10 & Topology of Autonomous Systems (AS) peering information inferred from Oregon route-views between March 31 2001 and May 26 2001 \cite{oregon} \tabularnewline
%        \assnapshot{} & 22963 & 48436 & 0.2304 & 32.9424 & 43.6526 & 11 & Snapshot of the structure of the Internet at the level of autonomous systems, reconstructed from BGP tables posted by the University of Oregon Route Views Project from data of July 22, 2006 \tabularnewline
        \hline
    \end{tabular}
  \begin{tablenotes}
    \item[1] $n$, $m$, $c$, $\sigma$, $s$, and $d$ are the number of nodes, the number of edges, clustering coefficient, standard deviation of degree distribution, skewness of degree distribution \cite{skewness}, and diameter of network, respectively
    \item[2] $d$ is calculated with the largest connected component if a network has multiple connected components
%        \vspace{-0.2in}
  \end{tablenotes}
  \end{threeparttable}
  \vspace{-0.9cm}
    \end{center}

    \end{table*}

\vspace{-0.1cm}
%!TEX root = main.tex
\section{Measuring Infection State} \label{sec:measure}
In this section, we describe the propagation properties of the SI epidemic that allows us to determine the unknown infection state of nodes. Then in Section~\ref{sec:IB} we introduce Infection Betweenness using the lessons learned in Section~\ref{sec:prop}.

\subsection{Propagation Properties}\label{sec:prop}
Under the assumption that an epidemic propagates from a single source to neighboring nodes following the SI model, we identify the following properties.
Let $S_{o}$ denote the set of observed susceptible nodes and $I_{o}$ be the set of observed infected nodes.
\begin{itemize}
    \item Property 1: If removing all nodes in  $S_{o}$ from the network disconnects the network, then one of the disconnected components contains all of the infected nodes
    \item Property 2: Let $S \in V$ be a cut set that divides $I_{o}$ into multiple components, then at least one node in $S$ is infected
\end{itemize}

    \begin{figure}[!hhh]
    \centering
        \includegraphics[scale=0.3]{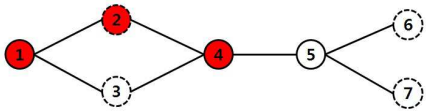}
    \caption{Red nodes and white nodes represent infected and susceptible nodes, respectively. Dotted circles (nodes 2, 3, 6, and 7) show nodes with unknown infection state and full circles (nodes 1, 4, and 5) show nodes with known infection state. From Property 1 we know that 6 and 7 are not infected. From Property 2 we know that either 2 or 3 must be infected.}
    \label{fig:example}
    \end{figure}

Consider the topology shown in Figure \ref{fig:example}. Removal of node 5, which is observed and susceptible divides the graph into two components, $\{1,2,3,4\}$ and $\{6\}$. Only the component $\{1,2,3,4\}$ contains infected nodes (Property 1). Since there is no propagation path from infected nodes without node 5, we can determine that nodes 6 and 7 are susceptible (deterministic susceptible nodes). Observed infected nodes $\{1,4\}$ divide into two components by removing nodes 2 and 3, which are not observed. Because the removal of nodes 2 and 3 places infected nodes 1 and 4 in distinct component, node 2 and/or 3 must be infected (Property 2).
%It is easy to show that these two properties always hold true. First, note that any propagation path between two infected nodes consists only of infected nodes; thus it cannot cross the cut consisting solely of susceptible nodes, which means that Property 1 holds true. Second, note that all infected nodes are in one connected component when an epidemic starts from a single source and infected nodes are not cured. If a cust divides the infected nodes into multiple components, at least one node in cut must be on a propagation path between components, that is, it has to be infected.
Using Property 1, we can reduce the number of nodes with unknown state by ignoring nodes in components that can be isolated by healthy nodes. In the rest of this paper, we focus on the reduced graph in which observed and deterministic susceptible nodes are excluded from the original graph. %In subsection \ref{subsec:deterministic}, we will see how many unobserved nodes in the considered networks become deterministic by Property 1.

Even though Property 2 does not provide a direct way for determining the unknown infection state, it points to the importance of a particular node in possibly infecting known infected nodes. Next, we use this insight to define a new centrality metric, Infection Betweenness.

%    \begin{figure}[!hhh]
%    \centering
%    \caption{Tree Example (The grey and white nodes are infected and susceptible nodes, respectively. Assume that the nodes with red circle are observed)}
%    \label{fig:example}
%        \includegraphics[scale=0.38]{./example.eps}
%    \end{figure}

\subsection{Infection Betweenness}\label{sec:IB}
%Here we introduce the Infection Betweenness ($I\!B$) that presents how likely an epidemic has to pass through a particular node. Based on this measure, we compute an approximate probability that an unobserved node appears in infected status.

Let $G'$ be a subgraph constructed by removing all nodes that must be healthy according to Property 1.
The number of paths of length $r\geq0$ between a pair of nodes $(i,j)$ in $G'$, $N_{ij}$, is
\begin{equation*}
    N^{(r)}_{ij} = \left( \mathbf{A}^r \right)_{ij},
\end{equation*}
where $\mathbf{A}$ is the adjacency matrix of $G'$.

Suppose that each path of length $r$ is given a weight $\alpha>0$; then
\begin{equation*}
    N_{ij} = \sum_{r=0}^{\infty} \alpha N^{(r)}_{ij} = \sum_{r=0}^{\infty} \left( \alpha^r \mathbf{A}^r \right)_{ij}.
\end{equation*}
is the weighted sum of paths from $i$ to $j$.
We can write this expression in matrix notation
\begin{equation*}
    \mathbf{N}=  \sum_{r=0}^{\infty}  \alpha ^r \mathbf{A}^r = (\mathbf{I} - \alpha \mathbf{A})^{-1}.
\end{equation*}

Let $N_u(i,j)$ denote the weighted sum of paths from node $i$ to $j$ through node $u$. Given $G'' = G' - \{u\}$, we can calculate $N_u(i,j)$ by subtracting the weighted sum of paths from $i$ to $j$ in $G''$ from the sum in $G'$; however, constructing $G''$ and performing the inverse operation for $\mathbf{N}$ of each $G''$ requires additional computation. Therefore, we resort to simple approximation $N_u(i,j) \approx \mathbf{N}_{iu} \times \mathbf{N}_{uj}$. Summing over all possible nodes $u \in V$ yields
\begin{equation*}
    \mathbf{M}_{ij} = \sum_{u \in V} N_u(i,j) = \sum_{u \in V} \mathbf{N}_{iu} \mathbf{N}_{uj} = (\mathbf{N}^2)_{ij}. \\
%    = \sum_{u \in V} \left(\sum_{r=0}^{\infty} \alpha ^r \mathbf{A}_{iu}^r\right) \left(\sum_{l=0}^{\infty} \alpha ^l \mathbf{A}_{uj}^l\right)   \\
%    = \sum_{u \in V} \sum_{h=0}^{\infty} \alpha ^h \sum_{r=0}^{h} \mathbf{A}_{iu}^r \mathbf{A}_{uj}^{h-r} \\
%    = \sum_{h=0}^{\infty} \alpha ^h \sum_{r=0}^{h} \mathbf{A}_{ij}^h = \sum_{h=0}^{\infty} \alpha ^h(h+1) \mathbf{A}_{ij}^h \\
%    = \dfrac{d}{d \alpha} \sum_{h=0}^{\infty} \alpha ^{h+1} \mathbf{A}_{ij}^h  \\
%   \mathbf{M} = \dfrac{d}{d \alpha} (\mathbf{I}/\alpha - \mathbf{A})^{-1}\\
%    = \dfrac{1}{\alpha ^2}\left( (\mathbf{I}/\alpha - \mathbf{A})^{-1} \right)^2 \\
\end{equation*}

We define the Infection Betweenness of node $u$ between two infected nodes $i$ and $j$ as:
\begin{equation*}
    B_u(i,j) = \dfrac{N_u(i,j)}{\mathbf{M}_{ij}},
\end{equation*}
which is the fraction of the weighted sum of path from $i$ to $j$ through $u$ over the total weighted sum of paths from $i$ to $j$; thus, node $u$ is more likely to be infected by node $i$ or $j$ as $B_u(i,j)$ increases.
If $B_u(i,j)$ is the probability that an infected node contaminates a neighbor then $1 - B_u(i,j)$ is approximately the probability that node $u$ was not infected when the infection traveled between nodes $i$ and $j$.
As a consequence, we approximate the probability that a node $u$ is infected as
\begin{equation} \label{eq:pIB}
    P(X_u = 1 | I_{o}) \approx 1 - \prod_{i,j \in I_{o},~i \neq j}\left( 1- B_u(i,j) \right) ,
\end{equation}
where $I_{o}$ is the set of observed infected nodes.

%!TEX root = main.tex
\section{Results} \label{sec:results}
%In this section we report the results of simulations to test our approach using Infection Betweenness based on machine learning algorithms.
%We also explore the impact of the amount of missing data as well as the impact of network characteristics on the performance of the algorithms.
%and the  can be coupled with machine learning approaches in order to provide the comparable performance with the {\em IB}-based estimation, having a lower computational overhead.
\vspace{-0.1cm}
\subsection{Setup}
We use datasets from several real world networks, which we classify into three categories: biological, collaboration, and device networks. In this paper, we use five datasets referred as to \yeast{} (biological), \power{}, \oregon{} (device), \grqc{}, and \hepth{} (collaboration). Table \ref{tab:topology} shows several characteristics of these networks, e.g., numbers of nodes and clustering coefficients. %We will see how network characteristics affect the performance of status estimation of unobserved nodes in the subsection \ref{subsec:impact_of_NC}.

We run batches of simulations for each network topology in Table~\ref{tab:topology} while varying the fraction of observed nodes from $5\%$ to $25\%$. In each run, we simulate a Susceptible-Infected (SI) cascade~\cite{Newman:Book} starting at a randomly selected seed node with infection rate $\lambda = 0.5$ until $10\%$ of nodes are infected. The parameter weight $\alpha$ of {\em IB} is set to 0.01 to guarantee to be less than the reciprocal of largest eigenvalue of adjacency matrix of the reduced graph (the condition that $\alpha$ must satisfy for the sum $\mathbf{N}$ in the equation of infection betweenness to converge). If a network has multiple connected components as does \yeast{}, we assume that an epidemic starts at a node in the largest connected component.
%The parameters that affect the performance are the observed fraction $\alpha$ and the ratio of malfunctioning nodes in observed nodes; for example, it is hard to predict status of others with an observed set which consists of only one node or some susceptible nodes. Our experimental goal is to show the proposed prediction scheme can be generally used for any type of network with any parameter setting; thus, our simulations are performed with the
In order to evaluate accuracy, we use three metrics: precision, recall, and F-Measure \cite{BOOK:DM}. %We note that nodes determined to be healthy by Property 1 are not counted towards our metrics.
\begin{itemize}
\item Precision: %the ratio of True Positives over the sum of True Positives and False Positives or
    the fraction of correctly classified nodes in nodes whose state is classified as infected%\footnote{True Positives and True Negative are the number of correctly identified nodes which are infected and susceptible, respectively, False Positives is the number of infected nodes falsely ascribed to susceptible status, and False Negatives is the number of susceptible nodes which are estimated as infected nodes.}~
\item Recall: %the ratio of True Positives over the sum of True Positives and False Negatives or
    the fraction of nodes whose state is classified as infected out of the all infected nodes
\item F-measure: a measure to consider both precision and recall in a single metric by taking their harmonic mean ($\frac{2 \times precision \times recall}{precision + recall}$).
\end{itemize}

\subsection{Incorporating Infection Betweenness into Machine Learning Algorithms}
In this section, we introduce a classification method using the feature based on infection betweenness and other node features based on machine learning (ML) algorithms.
%Classification is one of main tasks in the field of machine learning (ML). Our problem can be exactly reformulated to a classification problem of identifying the class to which unobserved node belong.
We choose three ML algorithms described in the following subsection. To apply these ML algorithms to experiments, we use the WEKA machine learning software suite \cite{WEKA}. For each topology, we collect the features of unobserved nodes from 30 simulation runs and then aggregate the collected feature instances into a training set. We run another 70 simulation runs to generate test data.

\subsubsection{Node features}
We consider six node characteristics that are available using information regarding network topology and the observed nodes, as features for building ML-based classifiers. The first five features are: degree normalized by the maximum degree in the network $D$, observed infected neighbor ratio $R$, betweenness centrality $C^{(b)}$, closeness centrality $C^{(c)}$, and eigenvector centrality $C^{(e)}$. We also include $P$ as a feature, defined as the Infection Betweenness probability that a node is infected shown in Eq.~\eqref{eq:pIB}.
\comment{
\begin{table}[h!!!]
\centering
\caption{Node Features}
\label{tab:feature}
\small
\begin{tabular}{cl}
  \hline
  Symbol & Explanation \\
  \hline
  \hline
  {\em IB} & Infection Betweenness \\
  $Deg$ & Node Degreee normalized by Maximum Degree \\
  $IN$ & Proportion of infected neighbors \\
  $BW$ & Betweenness Centrality \\
  $CL$ & Closeness Centrality \\
  $Eig$ & Eigenvector Centrality \\
  \hline
\end{tabular}
\end{table}
}
\subsubsection{Classifiers}\label{sec:cla}
{\bf Naive-Bayes}. Naive Bayes algorithms (NB, NBK) work under the assumption that there is no correlation between features given the class (infection state), \textbf{NB} derives a conditional probability for the relationships between the feature values and the class. To this end, \textbf{NB} must estimate the distribution of feature values.
For real-valued features, \textbf{NB} will assume that the values of each feature follows a particular distribution such as a Gaussian distribution. We evaluate the performance of \textbf{NB} to classify the state of unobserved nodes as well as Naive Bayes using and kernel density estimation (\textbf{NBK}); Kernel density estimation models use multiple (Gaussian) distributions, and generally provide more accurate results than using a single (Gaussian) distribution.

\textbf{C4.5 Decision Tree (C4.5)} constructs a decision tree model in which each internal node represents a test on features, each branch an outcome of the test, and each leaf node a class label \cite{C4.5}. In order to use a decision tree for classification, a given tuple (a set of feature values of a node), whose class we want to classify, walks through the decision tree from the root to a leaf. The label of final reached leaf is the classified class of which the tuple belong.

%The simplest way to combine results of various classifiers is to take a vote by averaging their probability estimates or numerical predictions. The problem with \emph{Voting} is that it is not clear which classifier is to be trusted. To overcome this limitation, \textbf{Stacked Generalization}, or \textbf{Stacking} for short, uses a \emph{metalearner} that replaces the voting procedure; \textbf{Stacking} exploits a \emph{metalearner} to discover which classifiers are most reliable. After learning with the training set and the results of other classifiers, the \emph{metalearner} makes a final decision based on the predictions of other classifiers~\cite{BOOK:DM}.

\subsubsection{Predictive Features}
\begin{figure}[h!!]
    \centering
        \subfigure[NB]{\includegraphics[scale=0.31]{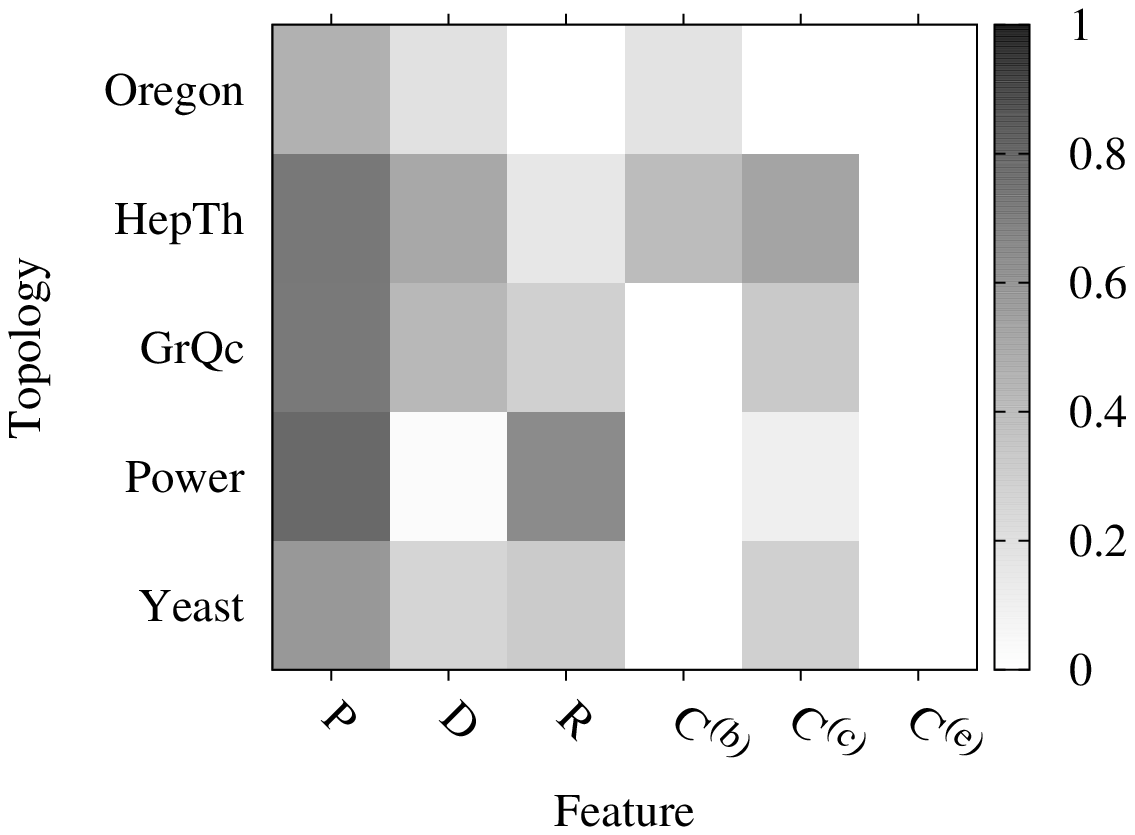} } %feature_c4_heatmap
        \subfigure[C4.5]{\includegraphics[scale=0.31]{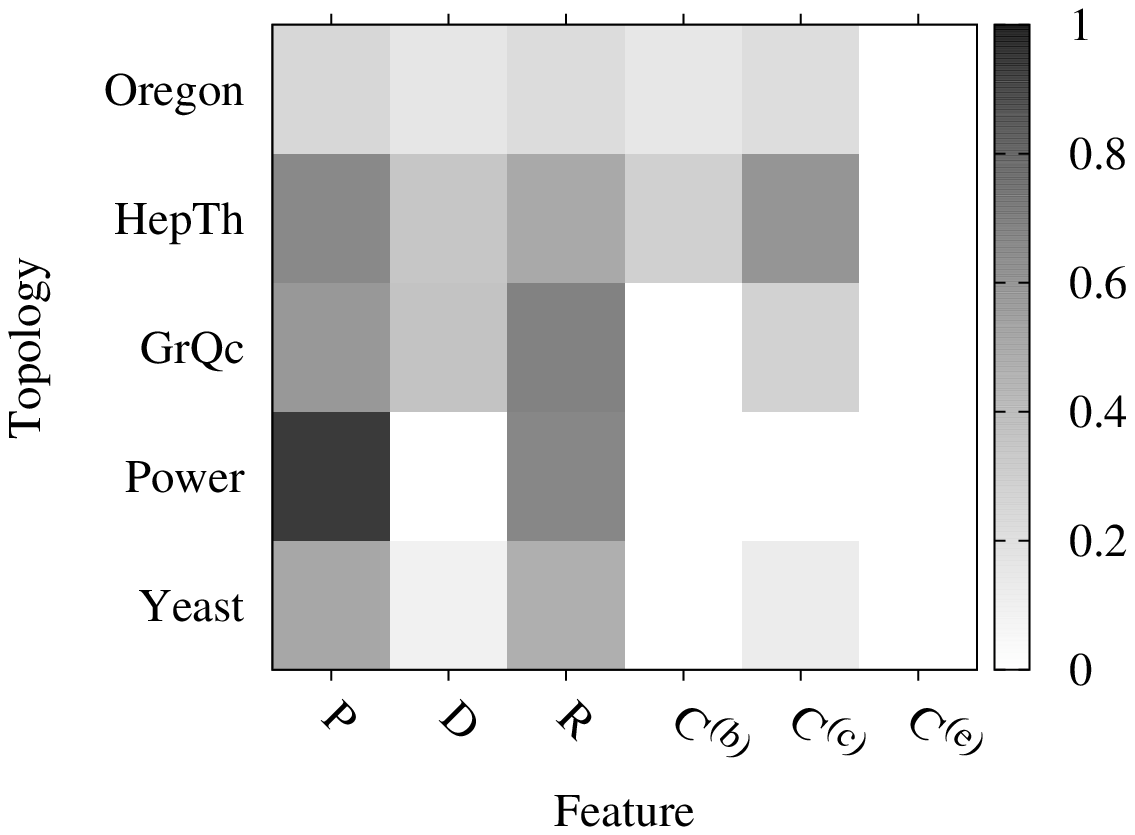} }
    \caption{Predictive power of each feature.}
    \label{fig:DiscPower}
    \vspace{-0.3cm}
    \end{figure}

In order to examine which features provides meaningful information for identifying latent infected nodes, we investigate the performance of ML-based classifiers with each feature. Figure \ref{fig:DiscPower} shows the average F-measure of \textbf{NB} and \textbf{C4.5} with each feature for all the networks.
The best feature will have F-measure close to one (darker squares).
We observe that the feature based on infection betweenness ($P$) produces the darkest column showing to be the best predictive feature in both \textbf{NB} and \textbf{C4.5} algorithms over nearly all networks.
In the case of \textbf{C4.5}, $R$ yields similar performance to $P$. We also see that $D$ and $C^{(c)}$ are also meaningful features in several networks although not as good as $P$. However, except for $P$, the effectiveness of other features differs significantly depending on the network and the ML algorithm. %In the next subsection, we examine ML-based classifiers that identify latent infected nodes by exploiting all these features.

\subsection{ML-based Infection State Prediction}
Above we found that the feature based on infection betweenness ($P$) to be the most predictive feature of a node's infection state.
In what follows, we show the accuracy of the ML-based classifiers over various scenarios when we incorporate all features (including $P$).

\subsubsection{Combining All Features}
\begin{figure}[h!!!]
    \centering
        \subfigure[F-measure when using only $P$]{ \includegraphics[scale=0.3]{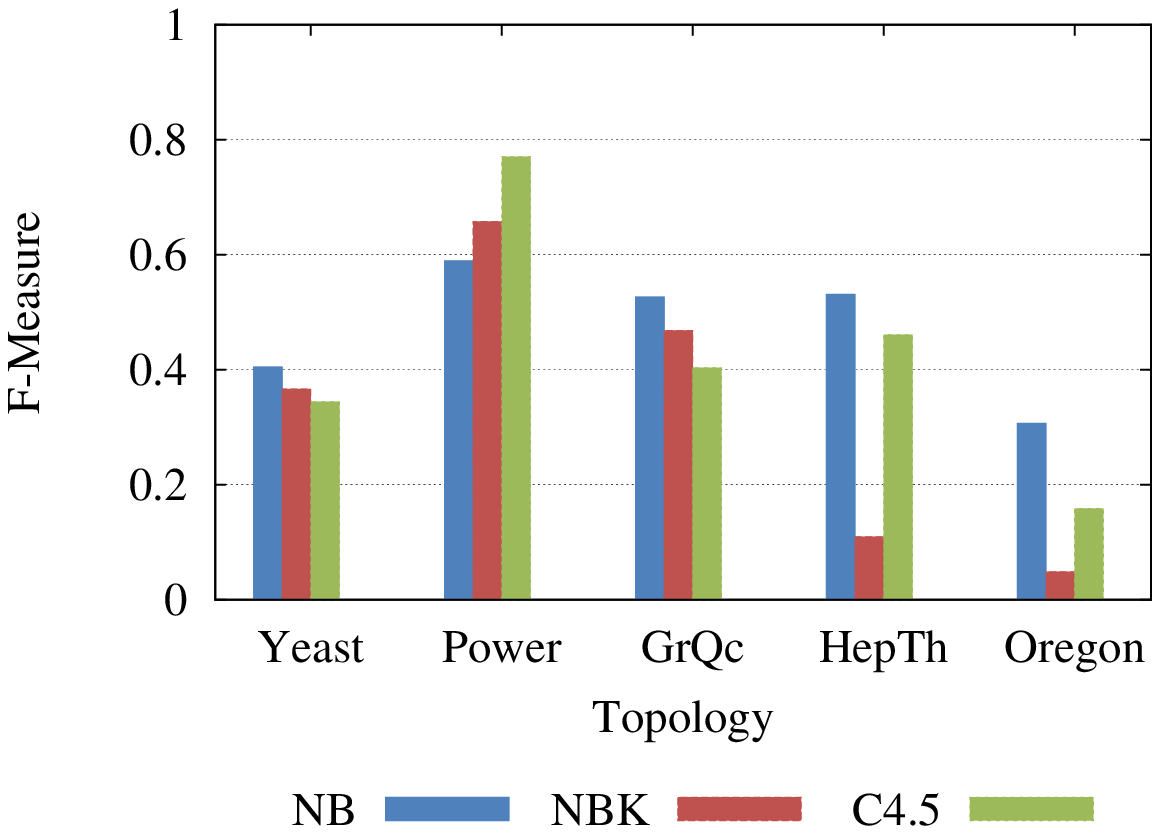}  }
        \subfigure[F-measure Enhancement by combining all features]{ \includegraphics[scale=0.3]{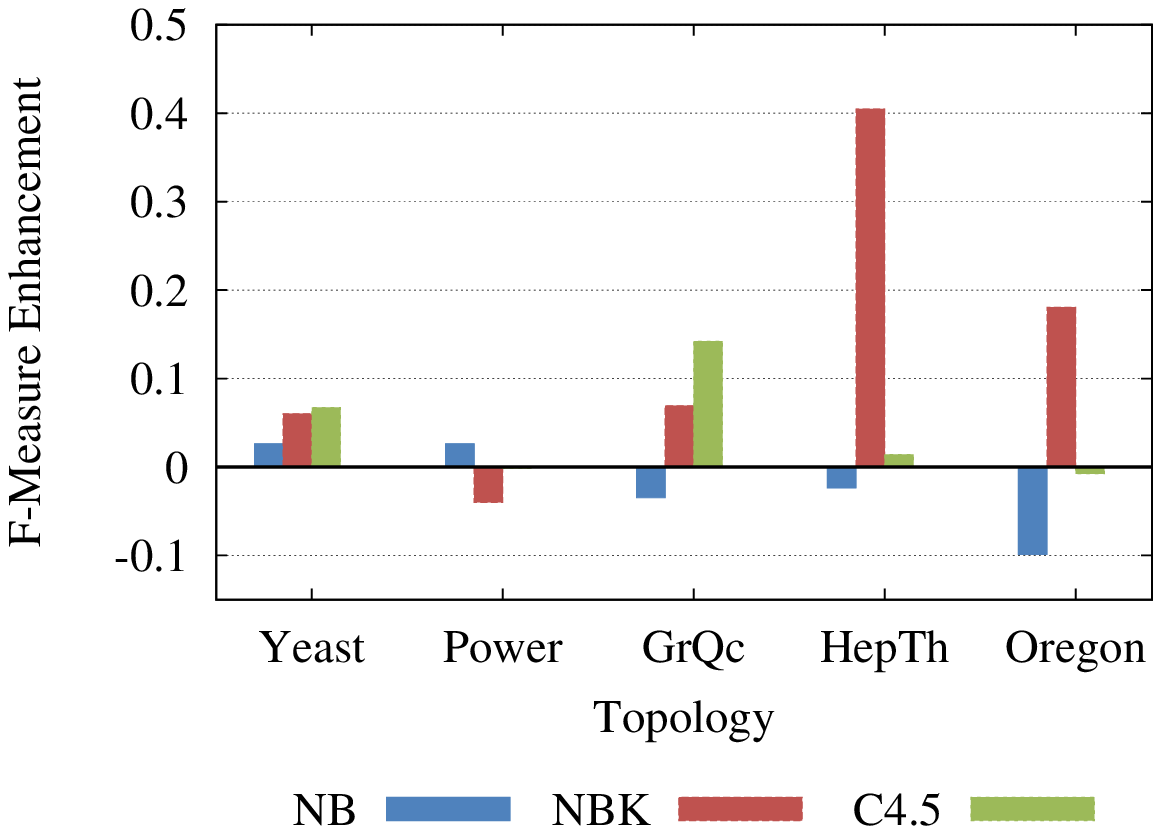}  }
    \caption{Performance of ML algorithms}
    \label{fig:exp_ML}
    \vspace{-0.2cm}
    \end{figure}

%We investigate the performance of each ML-based classifier using only $P$ or the all features after we create a cascade that infects approximately $10\%$ of the nodes in the network and then reveal the infection state of $15\%$ of the nodes (randomly chosen). 
In the following test we create cascades that infects approximately $10\%$ of the nodes in the network and then reveal the infection state of $15\%$ of the nodes (randomly selected).
Figure~\ref{fig:exp_ML}(a) shows the F-measure of each of Section~\ref{sec:cla} classifiers using only $P$ as feature. Note the significant performance difference between the classifiers for \hepth{} and \oregon{}; in these latter networks the F-measure of \textbf{NBK} is around 0.1, which is at least one third of that of the other classifiers. We also observe that both in \hepth{} and \oregon{} \textbf{NBK} using only $P$ is unable to correctly classifying most unknown infection states. Next, we compare the ML-based classifiers using all of the features to those using only $P$ in order to check whether more features can improve the performance of the classifiers by adding more features.
%Figure~\ref{fig:exp_ML}(b) shows the F-measure of each ML-based classifiers using all six features ({\em IB, Deg, IN, BW, CL, Eig}) minus the F-measure of {\bf IBalone}.
%Note that the IB alone in Eq.~\eqref{eq:pIB} allows for very simple classifier (which we denote {\bf IBraw}): if the value in Eq.~\eqref{eq:pIB} is greater than 0.5 we declare the node as infected, otherwise the node is healthy.
%Figure~\ref{fig:exp_ML} shows the F-measure of each ML-based classifiers using all six features ({\em IB, Deg, IN, BW, CL, Eig}) minus the F-measure of {\bf IBraw}.
%A good classifier should consistently have positive values.
%All the ML-based classifiers outperform {\bf IBraw} in \yeast{}, \grqc{}, and \hepth{}. Interestingly, {\bf IBraw} is better than the ML-based classifiers \textbf{NB} in \oregon{} and \power{} and \textbf{NBK} in \oregon{}, which points to the likely inappropriateness of the naive Bayes independence assumptions ({\bf IBraw} should never beat a classifier that contains the IB metric as a feature). The \textbf{C4.5} classifier consistently yields better performance over all the networks showing that it is worthy to boost the predictive power of {\em IB} with other node features.

    \begin{figure*}[ht!!!]
    \centering
        \subfigure[\yeast{}]{\includegraphics[scale=0.4]{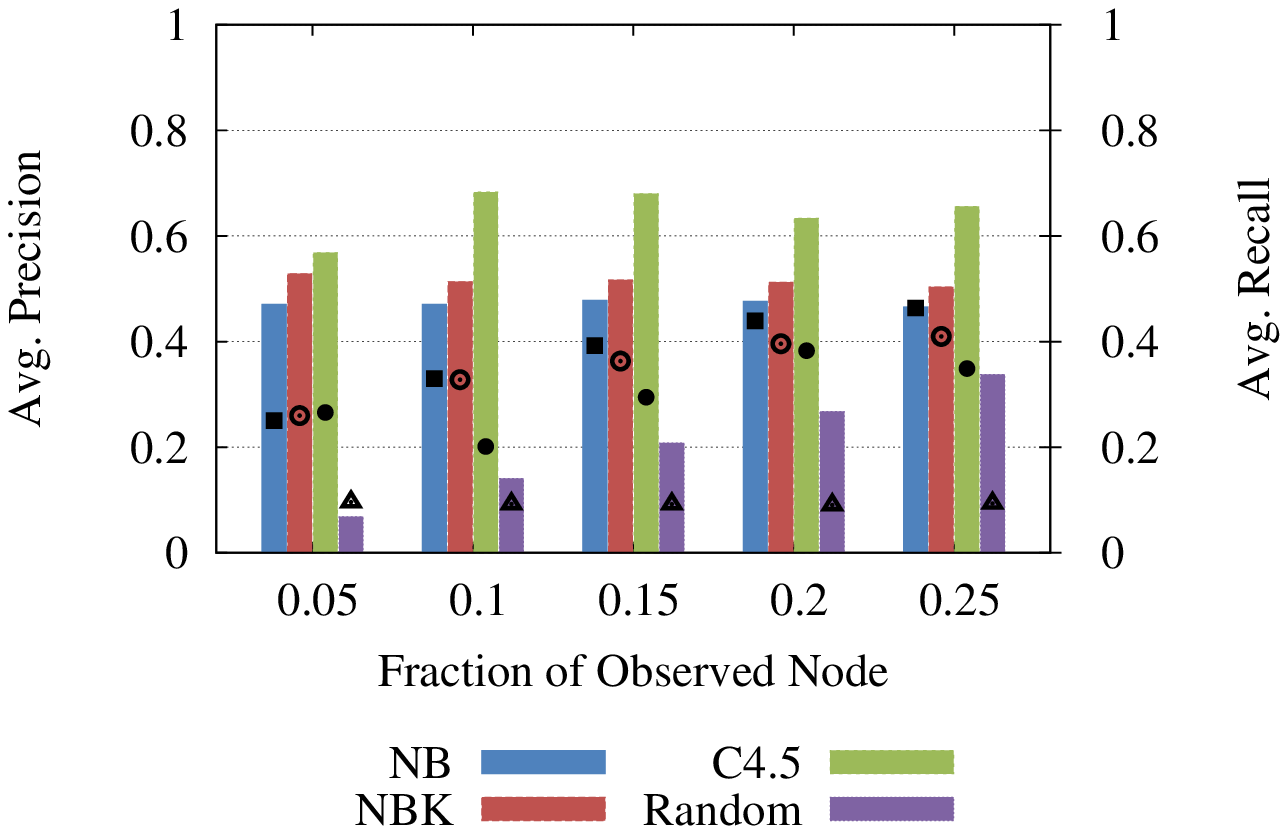}}
        \subfigure[\grqc{}]{\includegraphics[scale=0.4]{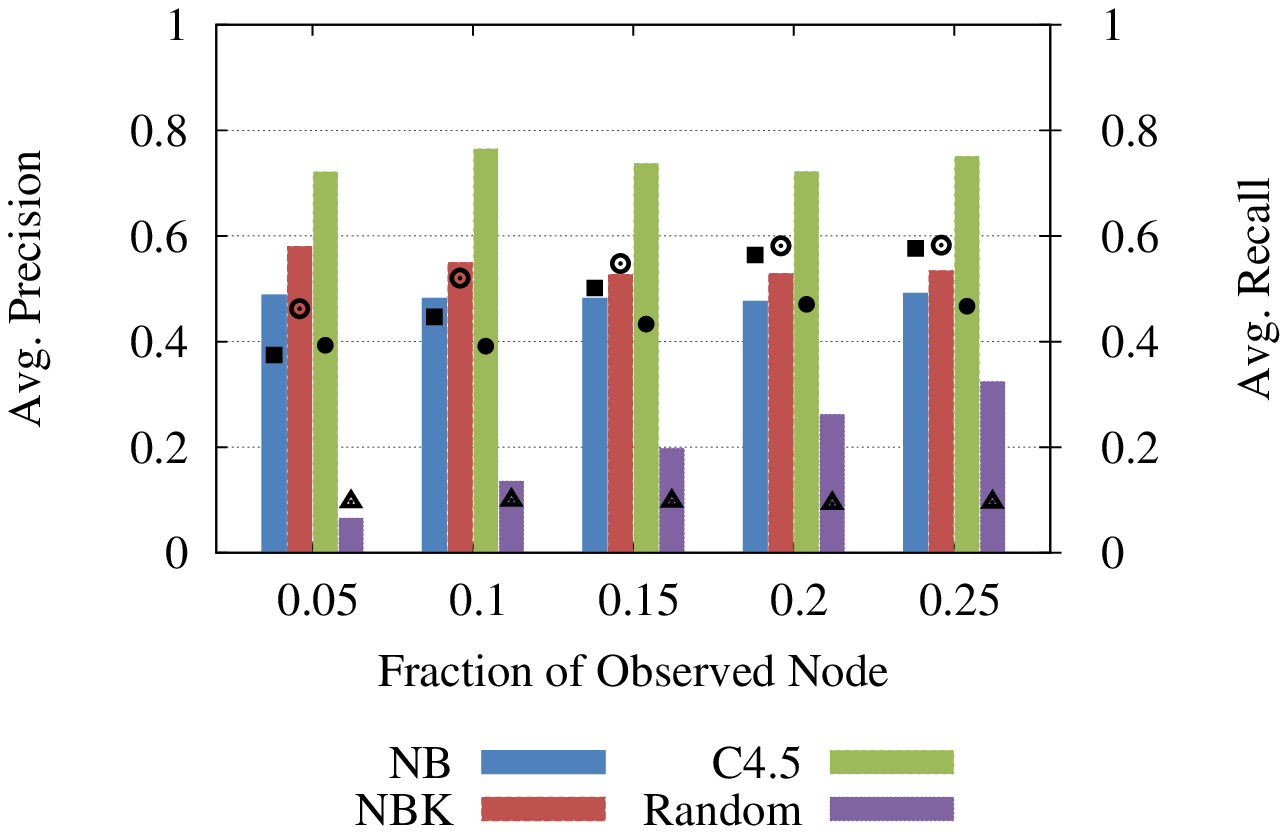}}
        \subfigure[\hepth{}]{\includegraphics[scale=0.4]{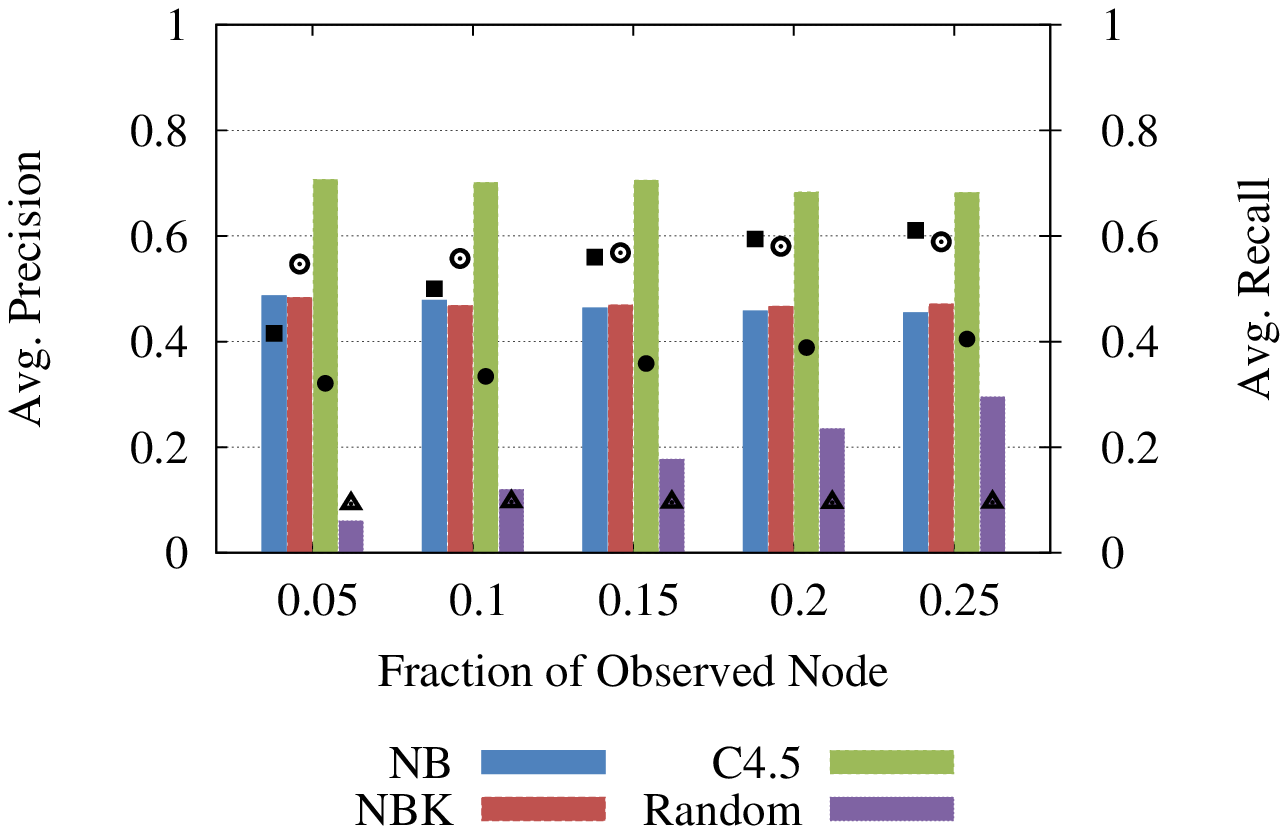}}
        \subfigure[\power{}]{\includegraphics[scale=0.4]{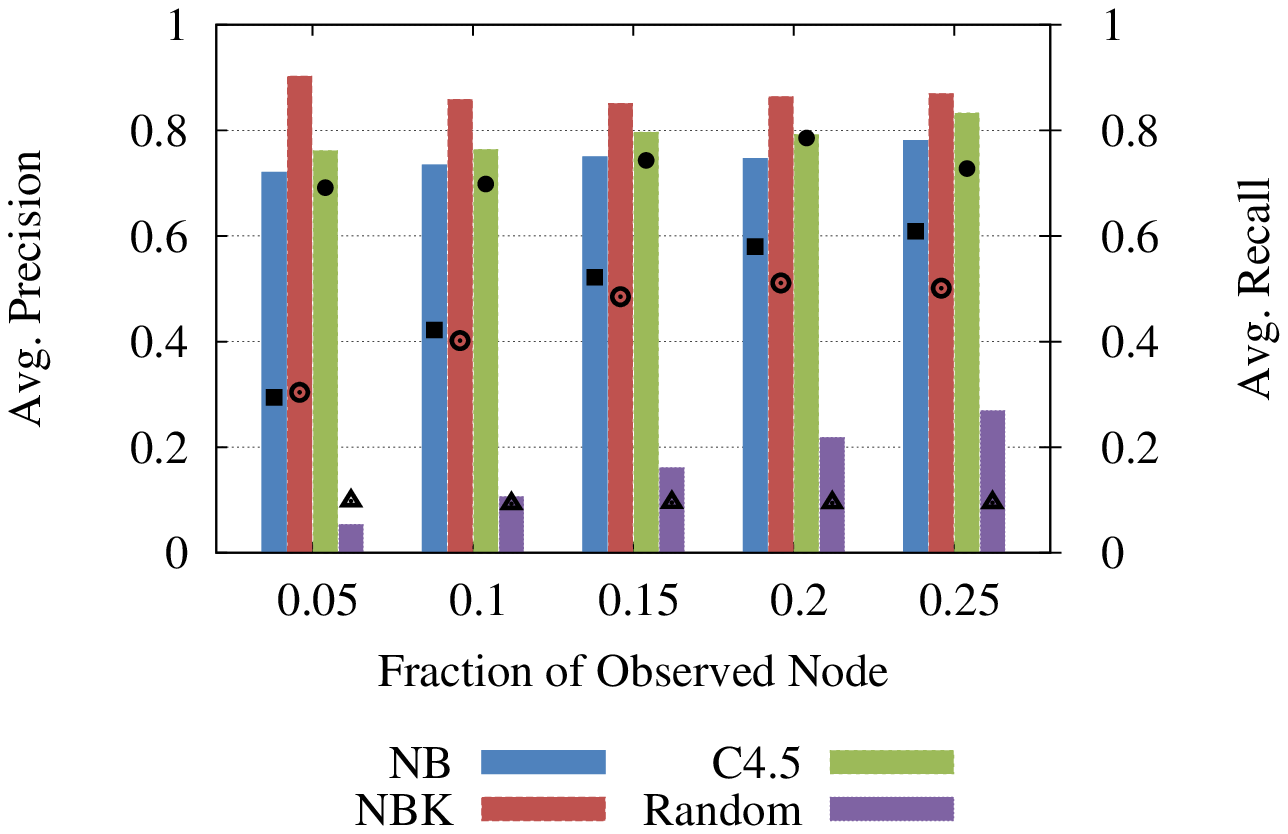}}
        \subfigure[\oregon{}]{\includegraphics[scale=0.4]{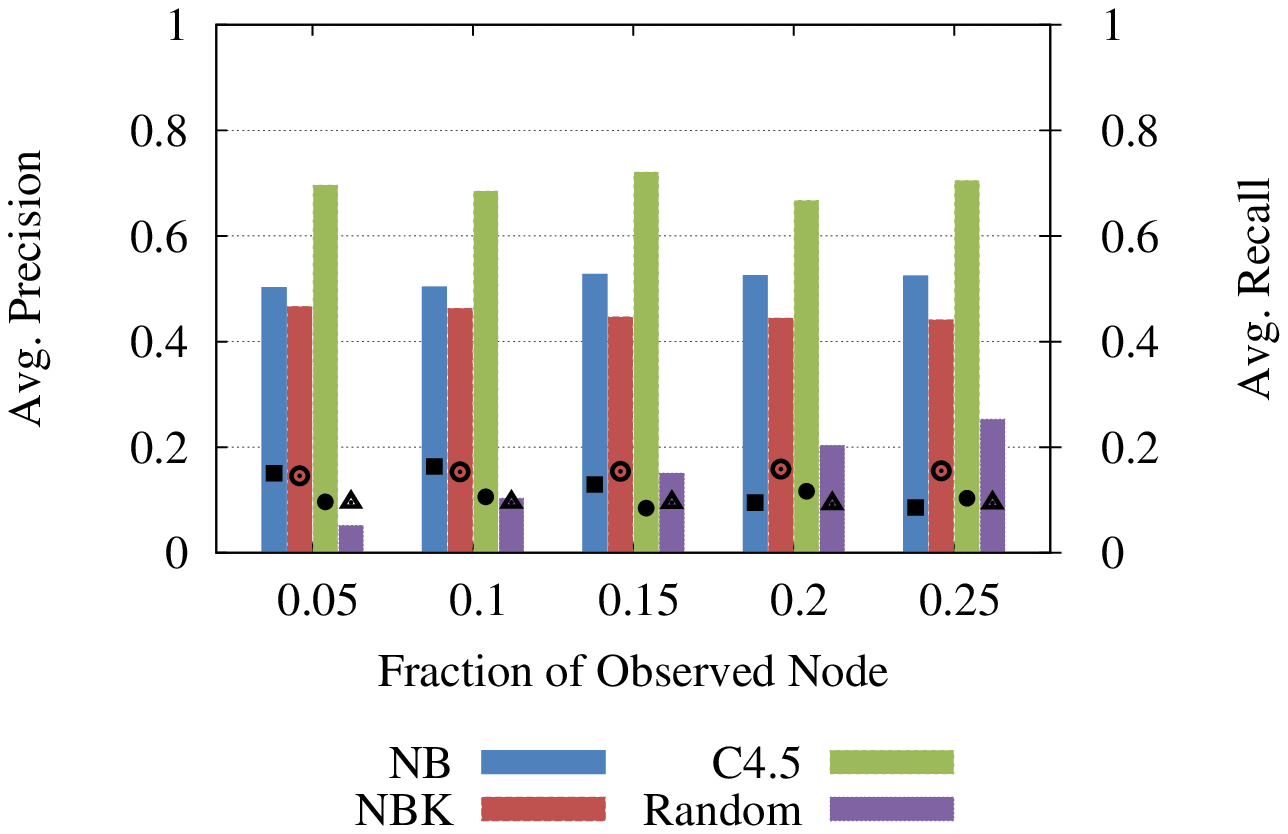}}
    \vspace{-0.2cm}
    \caption{Accuracy for varying fraction of nodes with observed state (Bar: Precision, Dot: Recall)}
    \label{fig:fo_precision}
    \vspace{-0.4cm}
    \end{figure*}

%Now, we compare the ML-based classifiers using all of the features to those using only $P$ (which we denote {\bf IBalone}).
Figure~\ref{fig:exp_ML}(b) shows the F-measure of each classifier using all six features (which includes $P$) minus the F-measure of the same classifiers using feature $P$alone. 
All classifiers using all features see performance improvements in \yeast{} compared to their single feature counterparts. 
In other networks by adding the extra five features the classifiers may, depending on the network, slightly underperform their $P$ feature counterparts, e.g., \textbf{NB} with only $P$ outperforms \textbf{NB} with $P$ and the extra five features in \grqc{}, \hepth{}, and \oregon{}. 
We conjecture that by adding more features (thus increasing the problem dimension) we make learning more difficulty, resulting in the observed performance degradation. 
For \textbf{C4.5} using all six features enhances performance in \yeast{} and \grqc{} while there is no significant change over other networks. 
We note then that  for \textbf{C4.5} feature $P$ is by far the most important feature as adding extra five features in most cases does little to increase the classification accuracy.
Note that except for \power{}, \textbf{NBK} with all features always yields better performance than \textbf{NBK} with feature $P$ alone: in particular, using all features increases the F-measure of \textbf{NBK} applied to  in \hepth{} and \oregon{} by around 0.4 and 0.2, respectively. Even in \power{}, the performance degradation of \textbf{NBK} by using all features is not noteworthy. 
It shows that we can improve the performance of a particular classifier by combining the infection betweenness feature $P$ with the other node features. 
In future work we will explore other classifiers such as classifiers based on random forests. %Interestingly, the {\bf C4.5} classifier is better than \textbf{NB} in \power{} and \grqc{} and \textbf{NBK} in \power{}, which points to the likely inappropriateness of the naive Bayes independence assumptions in those networks, while \textbf{NBK} is the most effective classifier in other networks.

\subsubsection{Prediction v.s.\ fraction of observed nodes} \label{subsec:impact_of}
In this section, we study the impact of the fraction of observed nodes on the accuracy of our classifiers with all six features.
%For this analysis, we ditch the excellent {\bf C4.5} classifier we built in favor of the simpler {\bf IBraw} as to better assess the impact of the fraction of observed nodes over the IB metric.
Figure~\ref{fig:fo_precision} compares the average precision and recall of each classifier according to the fraction of observed nodes. Again, the epidemic infects $10\%$ of nodes. Here, we also compare our classifiers against random-guessing (Random), which tosses a biased coin and with probability 0.1 (0.1 is the fraction of infected nodes) declares the node to be infected. As shown in Figure \ref{fig:fo_precision}, our classifiers outperform random-guessing both in precision and recall. 
Also, the precision and recall of our classifiers increases with the fraction of observed nodes; as expected, increasing the fraction of observed nodes provides more information about the infection state of the unobserved nodes.
%Note that the behavior of neighborhood heuristic is similar with the estimation that uses a $G\!I\!B$ which is normalized by the number of neighbors considering shortest paths only with the length of two.

In a closer look {\bf C4.5} exhibits the best precision over all classifiers on almost of all the networks: the only exception is \power{}, where \textbf{NBK} yields slightly better precision performance than \textbf{C4.5}. 
%These results show that the {\em IB} is more closely related to the probability that a node is infected than other measures.
%; it cannot achieve the equivalent precision to the neighborhood heuristic even with the high fraction of observed nodes.
Comparing the precisions of each network, we observe that our classifiers show the best precision in \power{} followed by \grqc{}, \hepth{}, \yeast{}, and \oregon{}. 
The power network is almost planar, likely making the classification task easier.
%Also, in the rest of our experiments, we see that the ML-based classifiers yield different behavior particularly in \power{}. 
In next section, we also explore which network characteristic affects on the performance of our classifiers.

We now look at the recall of each classifier. 
Figure \ref{fig:fo_precision} shows that \textbf{NBK} yields the best recall performance over all the networks except for \power{}. 
Note that the precisions of \textbf{NBK} is lower than that of \textbf{C4.5} except for \power{}. 
It means that \textbf{NBK} are more likely to find unknown infected nodes, but its classifications to the infected state are not as accurate as \textbf{C4.5}. 
 %As shown in Figure \ref{fig:fo_precision}, the results concerning recall exhibit similar trends to the results about precision; . In \power{}, the recall of classifying using {\em IB} alone (Eq.~\eqref{eq:pIB}) starts to decrease after $15\%$ of nodes are observed while the recall of neighborhood heuristic continues to increase; but the recall of classifying using {\em IB} alone is still larger than others. In particular, it becomes up to around $0.8$ when 10\% of nodes are observed. In other networks, {\em IB}-based estimations find $35\% \sim 50\%$ of infected nodes.
When considering each network, all classifiers yield better recall performance in \power{}. 
Also, \oregon{} remains the most difficult network to correctly classify the infected nodes. 
Even though all classifiers yield relatively high precisions (greater than 0.5) in \oregon{}, their recall performance in \oregon{} is less than 0.2, which is similar to that of random-guessing. 
It means that in \oregon{} our classifiers make correct decisions when they classify unknown states to infected, but many infected nodes are classified as healthy. 
In future work, we will explore a method to improve the recall performance of these classifiers using the estimated infected probability. 

\subsubsection{Impact of Network Characteristics}
\label{subsec:impact_of_NC}
We now investigate the impact of network characteristics on the performance of our classifiers (using all six features as before). 
To this end, we investigate the average correlation coefficient between the F-measure performance ranks and ranks of network characteristics for each network; for instance, \oregon{} has the highest Degree Skewness and \textbf{NBK} is the worst performer among the five classifiers: then we correlate 1 (highest rank in degree skewness) and 5 (worst classifier of five classifiers).
Table \ref{tab:correlation} presents the average Pearson's correlation coefficient~\cite{RANKCORR} between the ranks of network characteristics and F-measures.

\begin{table}[h!!!]
\vspace{-0.1cm}
\small
\centering
\caption{Correlation Coefficient between Ranks according to F-measure and Network Characteristics}
\label{tab:correlation}
\begin{tabular}{ccc}
  \hline
  \multirow{2}{*}{Characteristic} & Correlation  \\
  & \textbf{NB} & \textbf{NBK} \& \textbf{C4.5} \\
  \hline
  \hline
  Clustering Coefficient & 0.1 & 0.2 \\
  Standard Deviation of Degree & -0.7 & -0.6 \\
  Degree Skewness & -1.0 & -0.9 \\
  \hline
\end{tabular}
\vspace{-0.1cm}
\end{table}

As shown in Table \ref{tab:correlation}, the performance of the classifiers is strongly negatively correlated with degree skewness and the degree standard deviation. As the degree skewness and the degree standard deviation decrease the classifiers become more accurate. Interestingly, there is a little correlation between clustering coefficient and classification performance even though an epidemic is more likely to propagate to nodes in a same cluster. A validation with extensive experiment using more networks is part of our future work. %A theoretical analysis about the relationship between a degree skewness and correctness of {\em IB} is a subject for our futurework.

%!TEX root = main.tex
\section{Related work} \label{sec:relatedwork}
Several methods to detect the presence of network worms and rumor spreading nodes have been proposed in the literature. However, there has been little rigorous work done on inferring the infection state from incomplete data obtained at a relatively few observed nodes without the aid of infection timestamps.

Shah and Zaman~\cite{Shah} studied the problem of finding the source of a computer virus in a network. They focused on how to find the source among the set of infected nodes that are observed, which is different from our goal. Based on their metric called {\em rumor centrality}, they constructed a machine-learning estimator that finds the source exactly or within a few hops in networks. They also analyzed the asymptotic behavior of their virus source estimator for regular trees and geometric trees.

Sadikov et al.~\cite{Sadikov} present an estimation method of network properties, such as the number of weakly connected components, given a sampled network. By formulating a simple $k$-tree model and approximating it to the original network, their method can estimate the properties of original networks; they showed that their method can accurately estimate properties of the original network even when $90\%$ of nodes are not sampled.
Zou et al.~\cite{Zou:Worm} developed an early detection system to check the presence of a worm in the Internet. The proposed detection approach monitors traffic data at ingress/egree point of a local network. Even with the biased monitored data, it can accurately predict the overall vulnerable population size and estimate how many hosts are really infected in the global Internet system.

Closely related to our work is that of Gomez et al.~\cite{Gomez}, who develop an algorithm for inferring the network over which a diffusion propagates. Given the observed times when nodes become infected, they determine paths through which the diffusion most likely took, i.e., a directed graph where a contagion passed through. In contrast, our work tries to identify the infection state of each unobserved nodes given a limited number of nodes with known infection state and no infection timestamps.

%Sawaya et al.~\cite{Sawaya:DetectAttacker} proposed a flow-based attacker detection method focusing on the characteristics of attackers that send flows to both the object TCP port and generally closed TCP port in the global network. Thus, we need to inspect the flows from each node to identify whether it is from an attacker.
Jaikaeo et al.~\cite{Jaikaeo:DiagSensor} presented a malicious node detection method based on comparisons between neighboring nodes, performed on a central server. 
%Similar to the above approaches, 
It is not applicable without a central server which can directly access and inspect each node; thus, their method depends on being able to inspect each node individually.

\balance
\section{Conclusion} \label{sec:conclusion}
In this paper we studied how to identify the infected nodes without individually inspecting all nodes in the network. Based on the well known SI model, we defined the {\em Infection Betweenness} (IB) metric for identifying the latent infection status of nodes. Our empirical results show that the machine learning classifiers with the $I\!B$ metric as feature along with other network-wide features outperform random-guessing and the same classifiers without the $I\!B$ metric as a feature. We also analyzed the impact of the amount of missing data as well as the impact of network characteristics on the effectiveness of the algorithms.

Future work consists of performing more extensive experiments with larger networks, as well as a theoretical analysis about the relationship between network characteristics and performance of the algorithms.

\small
\bibliographystyle{abbrv}
\bibliography{refs}
\end{document}